\documentclass[]{article}
\usepackage{amsmath}
\usepackage{graphicx}
\usepackage{amssymb}
\begin{document}

\title{Tunable Band Structure Effects on  Ballistic Transport in Graphene Nanoribbons}

\author{O. Roslyak$^1$, Godfrey Gumbs$^{1,2}$ and Danhong Huang $^3$\\
$^1$Department of Physics and Astronomy,\\
Hunter College of  City University of New York,\\
695 Park Avenue, New York, NY 10065-50085 \\
$^2$ Donostia International Physics Center (DIPC),\\
P. de Manuel Lardizabal, 4, 20018 San Sebasti\'an,\\
Basque Country, Spain.\\
$^3$ Air Force Research Laboratory (AFRL/RVSS),\\ Kirtland Air
Force Base, NM 87117, USA}
\maketitle
\begin{abstract}
Graphene nanoribbons (GNR) in mutually perpendicular electric and magnetic fields
are shown to exhibit dramatic changes in their band structure and electron transport properties.
A strong electric field across the ribbon induces multiple chiral Dirac points,
closing the semiconducting gap in armchair GNR's. A perpendicular magnetic field induces
partially formed Landau levels as well as dispersive surface-bound states.
Each of the applied fields on its own preserves the even symmetry $E_{k} = E_{-k}$
of the subband dispersion. When applied together, they reverse the dispersion parity
to be odd and gives $E_{e,k} = -E_{h,-k}$ and mix the electron and hole subbands within
the  energy range corresponding to the change in potential  across the ribbon.
This  leads to oscillations of the ballistic conductance  within this energy range.
\end{abstract}

\par
Recent advantages in the fabrication techniques of graphene nanoribbons (GNR)
together with the long electron mean free path have stimulated considerable
interest in their potential applications as  interconnects in nano circuits.
Near the $K$ and $K^\prime$ Dirac points for infinite graphene, the electrons
are massless and chiral \cite{neto2009electronic}. The electronic properties of GNR are sensitive to the
geometry of their edges and the number of carbon atoms $N$ across the ribbon.
The GNR is thus classified as armchair (ANR), zigzag (ZNR) nanoribbons for even
$N$  and their counterpart anti-armchair (AANR), anti-zigzag (AZNR) for odd $N$.
The armchair confinement mixes $K$ and $K^\prime$ valleys creating chiral electrons
around the $\Gamma$ point. Chirality is the key ingredient for unimpeded electron
transport (Klein effect). Depending on $N$, modulo $3$, the ANR/AANR can be either
metallic or semiconducting making them suitable candidates for use as  field-
effect transistors. In contrast, the zigzag confinement does not mix the valleys
but rather intertwine their longitudinal and transverse momenta, creating edges-bound
quasi-particles between the $K$ and $K^\prime$ points. For ZNR/AZNR, the electrons
are not chiral (in the sense of projection of the pseudo-parity on the particle
momentum), and the electron transmission through a potential barrier is determined
by the electron pseudo-parity \cite{rainis2009andreev}. This quantity redefines the Klein effect as the
suppressed transmission through the barrier in ZNR, also known as the valley-valve effect \cite{roslyak2009klein}.
The latter is the basis for the proposed valley filters. The electron confinement in GNR
causes their properties to be quite sensitive to an applied electric \cite{novikov2007transverse,novikov2006energy,raza2008armchair}
or magnetic \cite{brey2006edge,ritter2008energy,perfetto2007quantum,golizadeh2008atomistic} field.
These changes are reflected in  measurable quantities
such as the ballistic conductivity and local density of states (LDOS) \cite{lyo2004quantized,li2009magnetoconductance}.
\par

\begin{figure}
\centering
  \includegraphics[width=0.45 \textwidth]{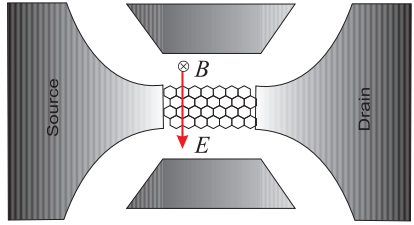}\\
  \caption{(Color online)  Schematic of an  ANR in the presence
   of an in-plane  electric field $E$ along the $y$ axis and a perpendicular
    magnetic  field $B$ along the $z$ axis.}
   \label{FIG:1}
\end{figure}

In this letter, we report on the individual and combined effects of an electric
 and magnetic field on the band structure and conductance of GNRs.
If only one of the fields is applied, it is well known that the time reversal symmetry
\footnote{Since we neglect spin, the action of the time reversal operator $\mathcal{T}$
amounts to reversing the direction of the wave vector propagation. The even/odd
 particle energy symmetry may be defined as $E_{n,k} = \pm \mathcal{T} E_{n,k} = \pm E_{n,-k}$}
of the energy bands for electrons and holes is preserved for all
 the types of GNRs we listed above. However, the combined effect of an electric
and magnetic field on the energy is to break the time reversal symmetry for both electrons and holes
and mix the energy bands. The effect of mixing on the differential conductance
and LDOS is presented below and our results are compared to those obtained
when only one of the two external fields is applied to an ANR with quantum
point contacts as illustrated schematically in Fig.\ \ref{FIG:1}. The ribbon
is attached to left (L) and right (R) leads serving as infinite electron reservoirs.
The R-lead is assumed to be the drain held at chemical potential $\mu$.
The L-lead is held at DC biased chemical potential $\mu+eV$  ($e$ is
the electron charge and $V$ is the bias potential) and serves
as the source. We choose coordinate axes so that the nanoribbon is along
the $x$ axis in the  $xy$-plane. Mutually perpendicular static electric
field $\mathcal{E}_y$ along the $y$ axis  and magnetic field $ \mathcal{B}_z $ along the $z$
axis are applied, as shown in Fig.\ \ref{FIG:1}.

We calculated the energy bands for graphene with sublattices A and B
in the tight-binding model \cite{lin2009ferromagnetism,neto2009electronic}. These are then separated into  hole
$\{{h}\}=\{{1 \le n < N}\}$ and electron $\{{e}\} = \{{N \le n < 2N}\}$
energy bands. The two component  wave function is a normalized $2N$
vector $\langle \Psi(x) \vert_{n,k} = \left({\langle \Psi_A (x)
\vert_{n,k},\langle \Psi_B (x) \vert_{n,k}}\right)$.
The electric field induces a potential across the ribbon
$U(y) = - e \mathcal{E}_y (y-W/2)=U_0 (y-W/2)$, where $W$ is the ribbon width.
The magnetic field modifies the wave vector as $k \rightarrow k - \mathcal{B}_z e y/c$,
which amounts to the Peierls phase in the hopping integrals \cite{liu2008orbital}.
The magnetic field  strength is assumed weak so we could take
the  energy levels as spin degenerate. The dispersion curves can be
experimentally observed via scanning tunneling microscopy (citation).
The tunneling current flowing through the microscope tip is proportional
to the LDOS given by

\begin{equation}
\label{EQ:LDOS}
\mathrm{LDOS}\left({E,x}\right) = \sum \limits_{n,k} \vert{\Psi_{n,k}(x)}\vert^2 \delta \left({E-E_{n,k}}\right)\ .
\end{equation}
The energy dispersion determines the ballistic
charge transport through the ribbon, at temperature $T$, by

\begin{gather}
\label{EQ:CURRENT}
Q(V,\mu,\mathcal{E}_y,\mathcal{B}_z,T) =\\
\notag
 = -\frac{2 e}{\hbar} \sum \limits_{n,k} \mathit{v}_{n,k}\left[{\theta(-\mathit{v}_{n,k}) f_{n,k}^>(1-f_{n,k}^<) + \theta(\mathit{v}_{n,k}) f_{n,k}^<(1-f_{n,k}^>)}\right]
\end{gather}
where $\mathit{v}_{n,k}=dE_{n,k}/d(\hbar k)$ is the carrier group velocity.
At $T=0$,  the Fermi function at the source contact is $f^{<}_{n,k} = 1-\theta \left({E_{n,k} - \mu - eV}\right)$ and for the drain, we have
$f^{>}_{n,k} = 1-\theta \left({E_{n,k} - \mu}\right)$. We note that
Eq.\ \eqref{EQ:CURRENT} does not assume any symmetry for the energy dispersion relation.
If the energy satisfies $E_{n,k} = E_{n,-k}$, we obtain the well-known
Landauer-Butikker formula \cite{buttiker1988absence}. That is the differential conductance
$G(\mu, \mathcal{E}_y,\mathcal{B}_z) = \left({\partial I/\partial V}\right)_{V=0}$ is determined by the number
of right-moving carriers through $\mathit{v}_{n,k}/\vert{\mathit{v}_{n,k}}\vert >0 $
at the chemical potential $E_{n,k}=\mu$. Alternatively, one may take
the difference between the local minima and maxima below the chemical potential
$E_{n,k}<\mu$ (citations).
\par

\begin{figure}[htbp]
\centering
\includegraphics[width=\textwidth]{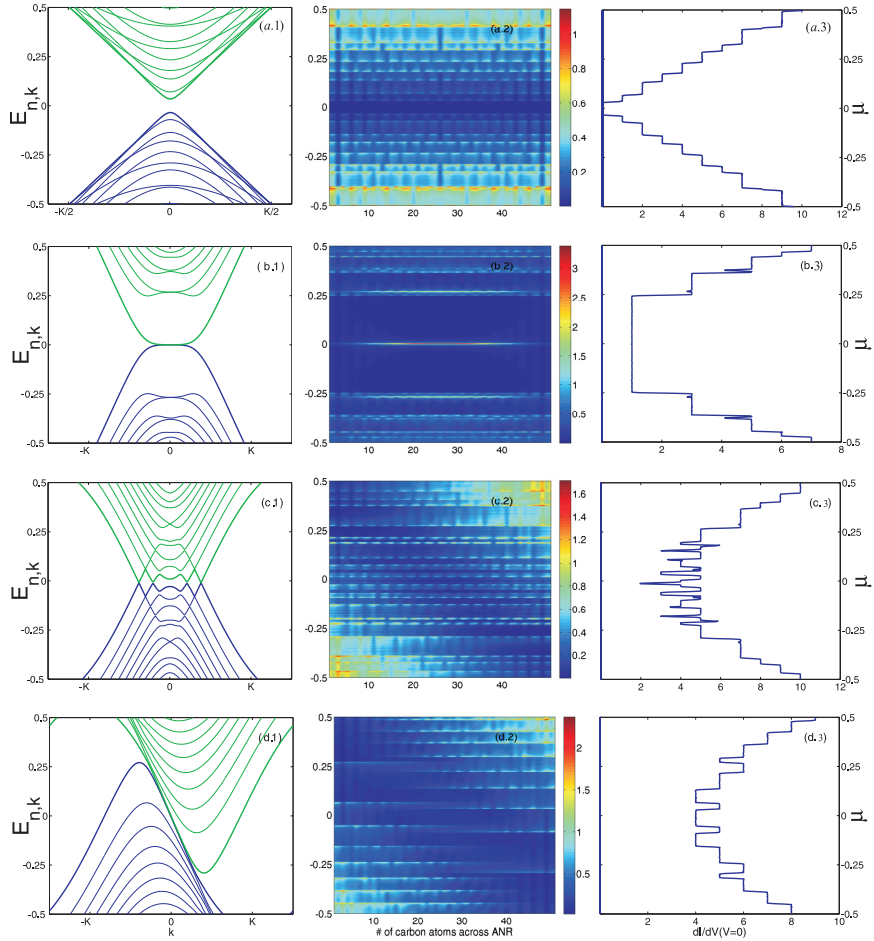}
\caption{(Color online). Panels (a) represent the dispersion curves for the electrons (green curves) in the conduction band and holes (blue curves) in the valence band. The lowest conduction and highest valence subbands are given by the thick curves. Panels (b) show local density of states. Panels (c) give the corresponding ballistic conductance in units of $2 e^2/h$. Panels (1) correspond to absence of the em. field. Panels (2) correspond to the sole magnetic field of the flux through a single hexagon placket $\phi/\phi_0 = 1/150$. Panel (3) show the effect of the sole electric field of the strength $U_0/t_0 = 1/2$. Panels (4) demonstrate the combined effects due to an electric and magnetic field  with the same strength as that employed   in panels (2)-(3).}
\label{FIG:2}
\end{figure}

Our numerical results for the energy bands, LDOS and  conductance
for  semiconducting ANR ($N=51$) in the presence of an  electric and/or magnetic
field are presented in Fig.\ref{FIG:2}. When either only an electric or magnetic field is applied $\mathcal{E}_y \mathcal{B}_z =0$,
the electron/hole energy bands  are symmetric with $E_{h,k}=-E_{e,k}$ and
time reversal symmetry is satisfied with $E_{n,k} = E_{n,-k}$  around the $k=0$ Dirac point
in Fig.\ref{FIG:2}(b.1). The latter means that if the time for the particle is
reversed, the particle retraces its path along the same electron/hole branch.
The LDOS also demonstrates the wave function symmetry with respect to the ribbon
center $\textrm{LDOS}(E,x)=\textrm{LDOS}(-E,x)=\textrm{LDOS}(E,-x)$.
In accordance with the Landauer-Butikker formalism, the conductivity demonstrates the familiar staggering pattern.
The magnetic field by itself distorts the weak dispersion ($n$ close to $N$) so that
the partially formed Landau levels $E_{n,0} \sim \sqrt{\mathcal{B}_z n}$ shows itself  up as
the flat parts in the dispersion curves. The lowest Landau level provides the single
conducting channel (along the ribbon edges), while the rest are doubly degenerate.
When the wave vector evolves from the Dirac point, the degeneracy is lifted and the
lowest subband acquires a local minimum.
Of these two effects,  the first one  can be observed in the LDOS, while the second
reveals itself as sharp spikes in the conductance as depicted in Fig.\ \ref{FIG:2}(b.3).
For the high energy subbands, when the radii of the Landau orbits
(spread of the wave function in Fig.\ \ref{FIG:2} (b.2)) become comparable with  the ribbon width,
the confinement effects dominate and the spectra become linear in magnetic field with $E_{n,0} \sim \mathcal{B}_z/n$.
These subbands are not degenerate.
\par

The main effect which the electric field has on the energy dispersion is
to fracture  Fermi surface into small pockets for $k \ne 0$, and thereby closing the semiconducting energy gap.
These zero energy points, where the group velocity
abruptly changes sign,  represent new Dirac points, which follows from the
chirality of the wave function in their vicinity \cite{brey2009emerging}. The rapid changes
in the group velocity cause the appearance of spikes in the conductance near
$\vert \mu \vert \le U_0/2$ and its step-like pattern is broken.
Due to the Dirac symmetry of the problem, the electron-hole band structure
remains  symmetric. The energy dispersion is not affected by magnetic field
at the original Dirac point $k=0$. Time reversal symmetry also persists.
The LDOS shows that at high energies the electric field confines the electrons
and holes near opposite boundaries. However, at low energies the LDOS does
not change across the ribbon, which is a manifestation of the Zitterbewegung
effect (attempt to confine Dirac fermions causes wave function delocalization \cite{neto2009electronic}).
With respect to the three cases considered above, we point out that the hallmark
of  Dirac fermions is the even symmetry of the dispersion with respect to the
wave vector, and steams from  time reversal symmetry.
Even though an attempt to confine them may lead to the broken electron/hole
symmetry \cite{peres2006dirac}, the wave vector symmetry still persists.
\par

We now turn our attention to the  most interesting case when both electric
and magnetic fields are applied together. Concurrent action of the electric field
dragging force, the Lorentz force and confinement by the ribbon edges  \emph{destroys}
 the Dirac symmetry of the problem so that $E_{n,k} \ne E_{n,-k}$ as shown in
Fig.\ref{FIG:2}(d.1). The dispersion distortion is different for the electrons
and holes, so the  symmetry between the conduction and valence bands is also broken.
On one hand,  the partially formed Landau levels get distorted by the confinement
 due to the electric field in conjunction with  the edges. Their degeneracy is
 also lifted. On the other hand, the magnetic field does not allow formation
 of additional Dirac points and wave function delocalization. At high energies,
 where the group velocity is decreased and the drag due to the electric field prevails.
The electrons and holes get gathered at the opposite ribbon edges (Fig.\ref{FIG:2}(d.2)).
For lower energies, in the region $\vert E_{n,k} \vert \le U_0/2$, the electron/hole dispersions overlap.
The electron bands have only local minima, whereas only the hole bands have
local maxima.  Regardless of the broken Dirac $k$ symmetry of the dispersion,
our numerical simulation of the differential conductivity shows that the Landauer-Butikker
expression still applies. Therefore, in the overlapping region
$\vert \mu \vert \le U_0/2$, the conductivity oscillates since the minimum of the
electron band is followed by the maximum on the hole band when the chemical
potential grows. As for possible applications of the broken Dirac symmetry,
the ribbon , subjected to mutually transverse electric and magnetic fields,
may serve as a field-effect transistor with a tunable working point.
An interesting feature of our results is that there is not only a breakdown in   the even-$k$
symmetry of the energy dispersion relation, but the energy bands are   reversed with  odd symmetry,
 satisfying $E_{e,k}=-E_{h,-k}$.   We explain this effect by adopting
 the method described in Ref. \ \cite{novikov2006energy,brey2009emerging}. Let us focus on the energy region
 close to the original Dirac point $k=0$,
 where the  unperturbed wave functions are governed by the conventional Dirac equation.
 Both applied fields are treated perturbatively. The effect of magnetic field is included
 through the wave vector replacement $k \rightarrow k - \mathcal{B}_z e y/c$. The electric field is
 treated by a chiral gauge transformation. This transformation shows that the spectrum at $k=0$
 is \emph{affected} by the electric field in the presence of the magnetic field. Regardless of the
 metallic or semiconducting ANR the electron and hole dispersion become degenerate around $k=0$ with
 $E_{e,k}=-E_{h,-k} \sim -\left({\mathcal{E}_y/\mathcal{B}_z}\right) k$.

\par

In conclusion, we have demonstrated that when GNRs are  placed in mutually perpendicular
electric and magnetic fields, there are dramatic changes in their band structure and
transport properties. The electric field across the ribbon induces multiple chiral Dirac points,
whereas a perpendicular magnetic field induces partially formed Landau levels accompanied by
dispersive surface-bound states. Each of the fields by itself preserves the original even parity
of the subband dispersion, i.e. $E_{n,k} = E_{n,-k}$, maintaining the Dirac fermion symmetry.
When applied together, their combined effect is to reverse  the dispersion parity to being
odd with $E_{e,k} = -E_{h,-k}$ and to mix electron and hole subbands within an energy range
equal to the potential drop across the ribbon.
Broken Dirac symmetry suppresses the wave function delocalization and the Zitterbewegung effect.
The Butikker formula for the conductance holds true for the odd $k$ symmetry.
This, in turn, causes the ballistic conductance to oscillate within this region which can be used
to design tunable field-effect transistors.

\par
This work was supported by contract \# FA 9453-07-C-0207 of AFRL
and the Air Force Office of Scientific Research
(AFOSR).

\bibliographystyle{plainnat}
\bibliography{Balistictransport}

\begin{thebibliography}{17}
\providecommand{\natexlab}[1]{#1}
\providecommand{\url}[1]{\texttt{#1}}
\expandafter\ifx\csname urlstyle\endcsname\relax
  \providecommand{\doi}[1]{doi: #1}\else
  \providecommand{\doi}{doi: \begingroup \urlstyle{rm}\Url}\fi

\bibitem[Brey and Fertig(2006)]{brey2006edge}
L.~Brey and HA~Fertig.
\newblock {Edge states and the quantized Hall effect in graphene}.
\newblock \emph{Physical Review B}, 73:\penalty0 195408, 2006.

\bibitem[Brey and Fertig(2009)]{brey2009emerging}
L.~Brey and H.A Fertig.
\newblock {Emerging zero modes for graphene in a periodic potential}.
\newblock \emph{Physical Review Letters}, 103:\penalty0 46809, 2009.

\bibitem[Buttiker(1988)]{buttiker1988absence}
M.~Buttiker.
\newblock {Absence of backscattering in the quantum Hall effect in multiprobe
  conductors}.
\newblock \emph{Physical Review B}, 38:\penalty0 9375, 1988.

\bibitem[Golizadeh-Mojarad et~al.(2008)Golizadeh-Mojarad, Zainuddin, Klimeck,
  and Datta]{golizadeh2008atomistic}
R.~Golizadeh-Mojarad, A.N.M. Zainuddin, G.~Klimeck, and S.~Datta.
\newblock {Atomistic non-equilibrium Green's function simulations of Graphene
  nano-ribbons in the quantum hall regime}.
\newblock \emph{Journal of Computational Electronics}, 7:\penalty0 407, 2008.

\bibitem[Li et~al.(2009)Li, Huang, Chang, Chang, and
  Lin]{li2009magnetoconductance}
T.S Li, Y.C Huang, S.C Chang, C.P Chang, and M.F Lin.
\newblock {Magnetoconductance of graphene nanoribbons}.
\newblock \emph{Philosophical Magazine}, 89:\penalty0 697, 2009.

\bibitem[Lin et~al.(2009)Lin, Hikihara, Jeng, Huang, Mou, and
  Hu]{lin2009ferromagnetism}
H.H. Lin, T.~Hikihara, H.T. Jeng, B.L. Huang, C.Y. Mou, and X.~Hu.
\newblock {Ferromagnetism in armchair graphene nanoribbons}.
\newblock \emph{Physical Review B}, 79:\penalty0 35405, 2009.

\bibitem[Liu et~al.(2008)Liu, Ma, Wright, and Zhang]{liu2008orbital}
J.~Liu, Z.~Ma, AR~Wright, and C.~Zhang.
\newblock {Orbital magnetization of graphene and graphene nanoribbons}.
\newblock \emph{Journal of Applied Physics}, 103:\penalty0 103711, 2008.

\bibitem[Lyo and Huang(2004)]{lyo2004quantized}
S.K Lyo and D.H Huang.
\newblock {Quantized magneto-thermopower in tunnel-coupled ballistic channels:
  sign reversal and oscillations}.
\newblock \emph{Journal of Physics Condensed Matter}, 16:\penalty0 3379, 2004.

\bibitem[Neto et~al.(2009)Neto, Guinea, Peres, Novoselov, and
  Geim]{neto2009electronic}
A.H.C. Neto, F.~Guinea, N.M.R Peres, K.S Novoselov, and A.K Geim.
\newblock {The electronic properties of graphene}.
\newblock \emph{Rev. Mod. Phys.}, 81:\penalty0 109, 2009.

\bibitem[Novikov(2007)]{novikov2007transverse}
D.S Novikov.
\newblock {Transverse field effect in graphene ribbons}.
\newblock \emph{Physical review letters}, 99:\penalty0 56802, 2007.

\bibitem[Novikov and Levitov(2006)]{novikov2006energy}
D.S Novikov and L.S Levitov.
\newblock {Energy Anomaly and Polarizability of Carbon Nanotubes}.
\newblock \emph{Physical review letters}, 96:\penalty0 36402, 2006.

\bibitem[Peres et~al.(2006)Peres, Castro~Neto, and Guinea]{peres2006dirac}
N.M.R Peres, A.H Castro~Neto, and F.~Guinea.
\newblock {Dirac fermion confinement in graphene}.
\newblock \emph{Physical Review B}, 73:\penalty0 241403, 2006.

\bibitem[Perfetto et~al.(2007)Perfetto, Gonz{\'a}lez, Guinea, Bellucci, and
  Onorato]{perfetto2007quantum}
E.~Perfetto, J.~Gonz{\'a}lez, F.~Guinea, S.~Bellucci, and P.~Onorato.
\newblock {Quantum Hall effect in carbon nanotubes and curved graphene strips}.
\newblock \emph{Physical Review B}, 76:\penalty0 125430, 2007.

\bibitem[Rainis et~al.(2009)Rainis, Taddei, Dolcini, Polini, and
  Fazio]{rainis2009andreev}
D.~Rainis, F.~Taddei, F.~Dolcini, M.~Polini, and R.~Fazio.
\newblock {Andreev reflection in graphene nanoribbons}.
\newblock \emph{Physical Review B}, 79:\penalty0 115131, 2009.

\bibitem[Raza and Kan(2008)]{raza2008armchair}
H.~Raza and E.C. Kan.
\newblock {Armchair graphene nanoribbons: Electronic structure and
  electric-field modulation}.
\newblock \emph{Physical Review B}, 77:\penalty0 245434, 2008.

\bibitem[Ritter et~al.(2008)Ritter, Makler, and Latge]{ritter2008energy}
C.~Ritter, S.S Makler, and A.~Latge.
\newblock {Energy-gap modulations of graphene ribbons under external fields: A
  theoretical study}.
\newblock \emph{Physical Review B}, 77:\penalty0 195443, 2008.

\bibitem[Roslyak et~al.(2009)Roslyak, Iurov, Gumbs, and
  Huang]{roslyak2009klein}
O.~Roslyak, A.~Iurov, G.~Gumbs, and D.~Huang.
\newblock {Klein Paradox in graphene nanoribbons}.
\newblock \emph{Submitted}, 2009.

\end{thebibliography}

\end{document}